\documentclass[aps,prb,twocolumn,showpacs,groupedaddress]{revtex4}
\usepackage{graphicx}

\begin{document}


\title{Spin current and shot noise from a quantum dot coupled to a quantized cavity field}

\author{Ivana Djuric and Chris P. Search}
\affiliation{Department of Physics and Engineering Physics,
Stevens Institute of Technology, Hoboken, NJ 07030}

\date{\today}

\begin{abstract}
We examine the spin current and the associated shot noise
generated in a quantum dot connected to normal leads with zero
bias voltage across the dot. The spin current is generated by spin
flip transitions induced by a quantized electromagnetic field
inside a cavity with one of the Zeeman states lying below the
Fermi level of the leads and the other above. In the limit of
strong Coulomb blockade, this model is analogous to the
Jaynes-Cummings model in quantum optics. We also calculate the
photon current and photon current shot noise resulting from
photons leaking out of the cavity. We show that the photon current
is equal to the spin current and that the spin current can be
significantly larger than for the case of a classical driving
field as a result of cavity losses. In addition to this, the
frequency dependent spin (photon) current shot noise show dips
(peaks) that are a result of the discrete nature of photons.

\end{abstract}

 \maketitle

\section{Introduction}
The field of spintronics has emerged as a new field in which the
spin degrees of freedom of charge carriers in solid state devices
are exploited for the purpose of information processing.
Manipulation of the spin degrees of freedom rather than the charge
has the advantage of longer coherence and relaxation times
\cite{zutic}. In contrast to spin-polarized charge currents
\cite{gijs}, pure spin currents,
$I_s=s(I_{\uparrow}-I_{\downarrow})$, are the result of an equal
number of spin up ($\uparrow$) and spin down ($\downarrow$) charge
carriers moving in the opposite direction so that the charge
current, $I_c=q(I_{\uparrow}+I_{\downarrow})$, is zero. Here,
$I_{\sigma}$ are the particle currents for each spin projection,
$s=\hbar/2$ the spin of the particles, and $q$ the charge of the
carriers.

There have been a number of proposals for generating spin currents
in semiconductor nanostructures, which include spin-orbit (SO)
dependent scattering off impurities (the extrinsic Spin-Hall
effect) \cite{extrinsic_SO}, and the intrinsic spin hall effect in
doped semiconductors \cite{intrinsic_SO} where the energy bands
are split due to spin orbit coupling \cite{rashba}. Interference
between one and two photon optical absorption in a semiconductor
can also be used to excite a pure spin current \cite{stevens}. A
quantum spin pump created by periodic shape deformations of an
open quantum dot \cite{mucciolo} has been proposed theoretically
and also demonstrated experimentally \cite{watson}. Quantum spin
pumps based on periodic variations of external potentials applied
to a quantum wire have also been proposed \cite{sharma}.
Additional proposals include quantum pumps that use independent
variations of localized magnetic fields in a two-dimensional
electron gas \cite{benjamin}, a quantum dot spin turnstile
\cite{blaauboer}, classical incoherent spin pumping \cite{sela},
and the use of superconducting leads \cite{xing}.

Electron spin resonance (ESR) in a quantum dot connected to leads
has been proposed as a way to generate a pure spin current when
there is a large Zeeman splitting \cite{wang-zhang,dong}. The spin
current and spin shot noise generated by such a quantum dot spin
battery has been analyzed by several groups
\cite{dong,sauret,wang2}. In previous work on this spin battery, a
classical electromagnetic (e.m.) field was used to induce the spin
flips. However, from quantum optics we know that the
Jaynes-Cummings model, which describes the interaction of a
two-level atom with a single mode of the quantized e.m. field
exhibits behavior not present when using a classical field
\cite{meystre}. This includes collapse and revivals in the
amplitude of Rabi oscillations caused by the discrete nature of
the photon number and the ability to create non-classical states
of the e.m. field as is done with the micromaser \cite{meystre}.

The intensity correlations of the e.m. field have been of central
importance in quantum optics for decades because they reveal
important information about the quantum state of the field that is
not present in the average intensity such as anti-bunching, a
purely non-classical effect, first observed in resonance
fluorescence \cite{kimble-RF}. More recently, the study of the
current shot noise, which is given by the current-current
correlator, has become the subject of intense theoretical study in
mesoscopic physics. This is because the shot noise contains
information about the quantum statistics of the charge carriers,
the interactions between particles, and the device structure that
is not present in measurements of the conductance \cite{Blanter}.
For example, the Pauli effect reduces the zero frequency charge
current noise below the Schottky value, $2e\langle I_c\rangle$,
corresponding to anti-bunching. However, it is difficult to
discriminate the effect of the Pauli principle and interparticle
interactions in the charge noise. Since the Pauli exclusion
principle does not effect fermions of opposite spin, the shot
noise in the spin current is a much more sensitive probe of the
interactions between particles \cite{sauret}. In many respects,
the current shot noise is the direct analogue of the second order
intensity correlations of the e.m. field.

Here we extend the model of the quantum dot spin battery
\cite{wang-zhang,dong} by studying for the first time the use of a
quantized cavity field to induce spin flips between Zeeman states.
We analyze the spin current and spin current shot noise and find
that the current produced by the quantum field can be larger than
that produced by a classical field. Moreover, the frequency
dependent spin shot noise shows unambiguous signatures of the
discrete nature of the photon states of the cavity.

In the section II, we develop our theoretical model in detail.
Section III discusses our numerical and analytic results for spin
current and shot noise. Finally, in section IV, we conclude with a
few comments on the experimental prospects for our work.

\begin{figure}[htb]
\includegraphics[height=3.5 in]{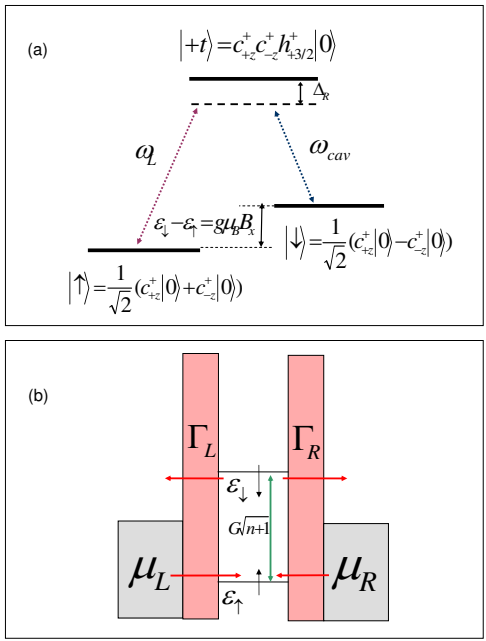}
\caption{(a) Raman transition between the electronic Zeeman
states, $|\uparrow,\downarrow \rangle$, via an intermediate trion
state, $|+t\rangle$, induced by a laser with frequency $\omega_L$
and a cavity mode with frequency $\omega_{cav}$. The spin
eigenstates along the direction of the magnetic field are
superpositions of spin eigenstates in the growth direction,
$\hat{c}^{\dagger}_{\pm z}|0\rangle$. (b) Diagram of quantum dot
indicating Zeeman energy levels in the dot and allowed tunnelling
between leads and dot. }\label{PIC.1}
\end{figure}

\section{Model}
For our quantum field spin battery, the quantum dot is embedded in
a high-Q microcavity. Strong coupling between an individual
quantum dot and a single mode of an optical microcavity has
recently been achieved\cite{reithmaier,peter}. Although electron
transport through quantum dots is most commonly studied using
electrically defined quantum dots in two dimensional electron
gases \cite{wiel}, these dots are not well suited for optical
experiments because of the strong optical absorption of the metal
electrodes used to define the dots \cite{strauf}. Self-assembled
quantum dots, as used in Ref. \onlinecite{reithmaier}, are readily
studied using optical techniques and in addition there have been
several experimental studies of electron transport including shot
noise through individual and coupled pairs of self-assembled InAs
quantum dots \cite{schmidt,ota,barthold}. Along a similar line,
the ability to control the tunnelling of electrons or holes
between self-assembled dots and a doped GaAs reservoir by a gate
voltage combined with simultaneous spectroscopic studies of these
charged quantum dots has been demonstrated \cite{petroff,atature}.
For the sake definiteness, we therefore assume that the quantum
dot under consideration is a self-assembled dot such as an InAs
dot embedded in GaAs.

There are two electron reservoirs at chemical potential, $\mu$,
that are coupled to the dot via tunnelling. (None of the results
presented here require two leads coupled to the dot. The only
difference between the two lead and one lead case being the total
rate at which electrons tunnel into or out of the dot since it is
the sum of the tunnelling rates from each lead.) Only a single
empty orbital energy level, $\epsilon_D$, of the dot lies close to
$\mu$. The Zeeman splitting between the two electron spin states
is $\Delta=\epsilon_{\downarrow}-\epsilon_{\uparrow}=g_x\mu_{B}B$
where $B$ is a static magnetic field along the x-axis that is
perpendicular to the growth direction (z). $\mu_B$ is the Bohr
magneton and $g_x$  is the electronic g-factor along the direction
of the magnetic field. The energy levels satisfy
$\epsilon_{\uparrow}=\epsilon_D-\Delta/2<\mu<\epsilon_{\downarrow}=\epsilon_D+\Delta/2$
so that only spin up electrons can tunnel into the dot and only
spin down electrons can tunnel out \cite{note}. In the limit of
infinite Coulomb Blockade that we consider, only a single electron
can occupy the dot resulting in the bare Hamiltonian for the dot,
$H_D=\epsilon_{\uparrow}\hat{c}^{\dagger}_{\uparrow}\hat{c}_{\uparrow}+\epsilon_{\downarrow}\hat{c}^{\dagger}_{\downarrow}\hat{c}_{\downarrow}$
where $\hat{c}_{\sigma}(\hat{c}^{\dagger}_{\sigma})$ are
annihilation (creation) operators for electrons in the dot with
spin $\sigma$ in the x-direction of the magnetic field.

Transitions between different spin states of the conduction band
electron in the dot are induced via a two-photon Raman transition
involving a strong laser field that may be treated classically and
a quantized mode of the microcavity similar to Ref.
\onlinecite{imamoglu}. The two optical fields couple the electron
spin states to a higher energy charged exciton state (known as a
trion) by creating an additional electron-hole pair in the dot
\cite{chen}. Recent experiments have shown how Raman scattering
via intermediate trion states can be used to generate electron
spin coherence \cite{greilich, dutt} and to pump the electron spin
into a specific spin state \cite{atature}.

The lowest energy trion states excited by $\sigma^{+}$ and
$\sigma^{-}$ polarized light consist of an electron singlet with a
heavy hole,
$|+t\rangle=\hat{c}^{\dagger}_{\uparrow}\hat{c}^{\dagger}_{\downarrow}\hat{h}^{\dagger}_{+3/2}|0\rangle$
and
$|-t\rangle=\hat{c}^{\dagger}_{\uparrow}\hat{c}^{\dagger}_{\downarrow}\hat{h}^{\dagger}_{-3/2}|0\rangle$
where $\hat{h}^{\dagger}_{\pm 3/2}$ are heavy hole creation
operators with spin projections $\pm \hbar 3/2$ along the z-axis
and $|0\rangle$ is the empty dot state. The $\sigma^{+}$ polarized
laser with frequency $\omega_l$ and Rabi frequency $\Omega_l$
couples each of the electron spin states to the $|+t\rangle$ trion
state. On the other hand, the x-polarized cavity field with vacuum
Rabi frequency $g_{cav}$ and frequency $\omega_c$ couples the spin
states to both the $|+t\rangle$ and $|-t\rangle$ states. When the
two fields are far detuned from the creation energy for the
trions, the intermediate trion states can be adiabatically
eliminated to give the Hamiltonian,
\begin{equation}
H_{cav}=\hbar\omega_c\hat{a}^{\dagger}\hat{a}+G(\hat{a}^{\dagger}\hat{c}^{\dagger}_{\downarrow}\hat{c}_{\uparrow}e^{-i\omega_l
t}+h.c.).
\end{equation}
where $\hat{a}$ is a bosonic annihilation operator for the cavity
field. Here $G=g_{cav}\Omega_l/4\Delta_R$ and $\Delta_R$ is the
detuning of the laser and cavity mode from the $|+t\rangle$
creation energy. AC stark shifts of the electron energy levels due
to the optical fields have been absorbed into a redefinition of
$\epsilon_{\sigma}$.

By transforming to a rotating frame for the electron operators,
$\hat{c}_{\uparrow}=\hat{C}_{\uparrow}\exp(i\omega_lt/2)$ and
$\hat{c}_{\downarrow}=\hat{C}_{\downarrow}\exp(-i\omega_lt/2)$ the
explicit time dependence is removed from $H_{cav}$ and $H_D$
becomes,
\begin{equation}
H'_D=\epsilon_D(\hat{C}^{\dagger}_{\uparrow}\hat{C}_{\uparrow}+\hat{C}^{\dagger}_{\downarrow}\hat{C}_{\downarrow})+
(\Delta-\omega_l)(\hat{C}^{\dagger}_{\downarrow}\hat{C}_{\downarrow}-\hat{C}^{\dagger}_{\uparrow}\hat{C}_{\uparrow})/2.
\end{equation}
Since $\Delta< \omega_l$ for optical frequencies and typical
Zeeman splittings, the energies of the spin states are inverted in
the rotating frame. The two-photon resonance can then be seen to
be $\omega_{cav}=\omega_l-\Delta$. This level inversion coincides
with $H_{cav}$ where one sees that the $|\uparrow\rangle
\rightarrow |\downarrow\rangle$ transition creates photons. Energy
is in fact conserved in this process because the energy for the
spin flip and the cavity photon comes from the laser field, which
is treated in the undepleted pump approximation. $H_{cav}+H'_D$ is
the Jaynes-Cummings Hamiltonian, which results in independent time
evolution for each of the two-state manifolds $\{|\uparrow,n
\rangle, |\downarrow,n+1 \rangle\}$ characterized by the photon
number $n$. Coupling between these manifolds is a result of the
dot-lead coupling and the cavity damping.

The coupling between the lead and the dot can be treated using a
master equation similar to the one developed in Ref.
\onlinecite{dong-PRB}. We define the matrix elements of the
dot-cavity density operator to be
$\rho^{(n,m)}_{\sigma,\sigma'}=\langle
n,\sigma|\hat{\rho}|\sigma',m\rangle$ where $|\sigma,n\rangle$
represents a state with $n$ photons in the cavity and
$\sigma=0,\uparrow,\downarrow$ corresponding to no electrons, one
spin up, or one spin down electron, respectively. The specific
form of the master equations for the lead coupling are
\begin{eqnarray}
\dot{\rho}^{(n,m)}_{0,0}|_{lead}&=& \Gamma^{(-)}_{\downarrow}
\rho^{(n,m)}_{\downarrow,\downarrow}-\Gamma^{(+)}_\uparrow\rho^{(n,m)}_{0,0}
\\
\dot{\rho}^{(n,m)}_{\uparrow,\uparrow}|_{lead}&=&\Gamma^{(+)}_{\uparrow}\rho_{0,0}^{(n,m)}
\\
\dot{\rho}^{(n,m)}_{\downarrow,\downarrow}|_{lead}&=&-\Gamma_{\downarrow}^{(-)}\rho^{(n,m)}_{\downarrow,\downarrow}
\\
\dot{\rho}^{(n,m)}_{\uparrow,\downarrow}|_{lead}&=&-\Gamma^{(-)}_{\downarrow}\rho^{(n,m)}_{\uparrow,\downarrow}/2.
\end{eqnarray}
The rate at which spin up electrons tunnel into the dot is given
by
$\Gamma^{(+)}_{\uparrow}=\sum_{\eta}\Gamma^{(+)}_{\uparrow,\eta}=2\pi\sum_{\eta}\sum_{k}|t_{\eta,k,\uparrow}|^2\delta(\omega-\epsilon_{\eta,k,\uparrow})f_{\eta}(\epsilon_{\uparrow})$
where $f_{\eta}(\omega)$ is the Fermi distribution for the leads
and $t_{\eta,k,\sigma}$ is the tunnelling amplitude for an
electron from lead $\eta$ with momentum $\hbar k$, spin $\sigma$,
and energy $\epsilon_{\eta,k,\sigma}$.
$\Gamma^{(-)}_{\downarrow}=\sum_{\eta}\Gamma^{(-)}_{\downarrow,\eta}=2\pi\sum_{\eta}\sum_{k}|t_{\eta,k,\downarrow}|^2\delta(\omega-\epsilon_{\eta,k,\downarrow})(1-f_{\eta}(\epsilon_{\downarrow}))$
is the rate at which spin down electrons tunnel out of the dot.

Since the pump laser interacting with the quantum dot is a source
of energy for the cavity field, the photons in the cavity field
would increase without bound and not reach a steady state in the
absence of cavity damping. In order to describe the damping of the
cavity we use a zero temperature ($k_BT\ll \hbar\omega_{cav}$)
Born-Markov master equation with the matrix elements
\cite{meystre},
\begin{eqnarray}
\dot{\rho}^{(n,m)}_{\sigma,\sigma'}|_{cavity}&=&-\Gamma_{cav}(n+m)\rho^{(n,m)}_{\sigma,\sigma'}/2
\nonumber
\\&+&\Gamma_{cav}\sqrt{(n+1)(m+1)}\rho^{(n+1,m+1)}_{\sigma,\sigma'}.
\end{eqnarray}
Lastly, the unitary time evolution between the dot and the cavity
field is given by
$\dot{\rho}^{(n,m)}_{\sigma,\sigma'}|_{d-c}=(i\hbar)^{-1}\langle
n,\sigma|[H'_D+H_{cav},\hat{\rho}]|m,\sigma'\rangle$. The complete
master equation for the dot-cavity system is then given by
\begin{equation}
\dot{\rho}^{(n,m)}_{\sigma,\sigma'}=\dot{\rho}^{(n,m)}_{\sigma,\sigma'}|_{lead}+
\dot{\rho}^{(n,m)}_{\sigma,\sigma'}|_{cavity}+\dot{\rho}^{(n,m)}_{\sigma,\sigma'}|_{d-c}.\label{master}\end{equation}
For simplicity, we consider only two photon resonance,
$\omega_{cav}=\omega_l-\Delta$, which is easily achieved by
properly tuning the laser frequency.

We also assume that the couple between the left and right leads
and the dot are the same and that the tunnelling between the leads
and the dot is spin independent, $\Gamma^{(+)}_{\uparrow
L}=\Gamma^{(-)}_{\downarrow L}=\Gamma^{(+)}_{\uparrow
R}=\Gamma^{(-)}_{\downarrow R}=\Gamma$, where we will use $\Gamma$
as our unit of energy from here on. Although there have been no
measurements of the spin dependence of tunnelling in self
assembled dots, the tunnelling rates in electrically defined
quantum dots can be spin dependent \cite{elzerman,hanson}.
However, our numerical results indicate that as long as
$\Gamma^{(+)}_{\uparrow L}+\Gamma^{(+)}_{\uparrow R}$ is similar
in size to $\Gamma^{(-)}_{\downarrow L}+\Gamma^{(-)}_{\downarrow
R}$, the results presented here for equal spin tunnelling rates
will show no significant difference from case of unequal
tunnelling rates.

\section{Results}
Due to the identical coupling to both leads, the currents will be
the same in both leads, $I_{L,\sigma}=I_{R,\sigma}=I_{\sigma}$
where $I_{\eta,\uparrow(\downarrow)}$ is the spin-up(down)
electron particle current in the lead $\eta=L,R$. The average spin
current, $I_{s}=s(I_{\eta,\uparrow}-I_{\eta,\downarrow})$, is then
independent of the lead with the stationary currents given by
$I_{\uparrow}=\Gamma\bar{\rho}_{0,0}$ and
$I_{\downarrow}=-\Gamma\bar{\rho}_{\downarrow,\downarrow}$
\cite{ivana}. Here, the over bar denotes the steady state solution
and in particular
$\bar{\rho}_{i,i}=\sum_{n}\bar{\rho}^{(n,n)}_{i,i}$ is the steady
state solution of Eq. \ref{master} traced over the state of the
cavity. It is easy to show that
$\bar{\rho}_{\downarrow,\downarrow}=\bar{\rho}_{0,0}$. It then
follows from Eq. (\ref{master}) that the spin current can be
expressed as
\begin{equation}
I_{s}=2s\Gamma\bar{\rho}_{\downarrow,\downarrow}=
siG\sum_{n}\sqrt{n}(\bar{\rho}^{(n,n-1)}_{\downarrow,\uparrow}-\bar{\rho}^{(n-1,n)}_{\uparrow,\downarrow})
\label{I_S}
\end{equation}
The photon current leaving the cavity is given by
$I_{photon}=\Gamma_{cav}\langle n_{cav}\rangle$ where $\langle
n_{cav}\rangle=\sum_{\sigma,n}n\bar{\rho}^{n,n}_{\sigma,\sigma}$
is the average number of photons inside the cavity. In the steady
state one finds using Eq. \ref{I_S} that $I_{photon}=I_{s}/s$.

\begin{figure}[htb]
\includegraphics[height=1.8 in,width=3.in]{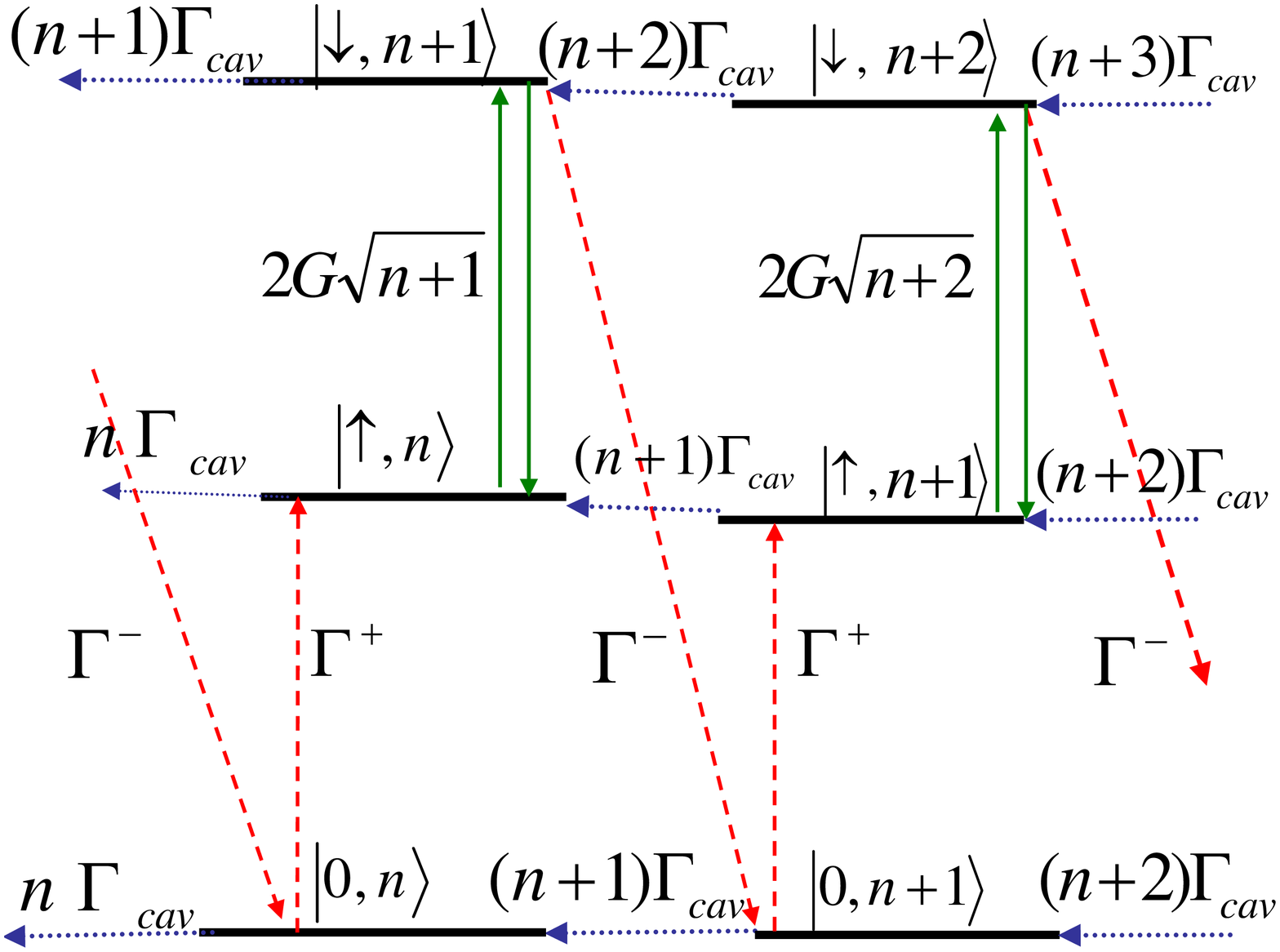}
\caption{Transitions between states $|\sigma,n\rangle$ with the
rates for each transition. Here $\sigma=0,\uparrow,\downarrow$ is
the dot state and $n$ the photon number in the cavity
}\label{PIC.1}
\end{figure}

The equality of the spin and photo-current is because exactly one
photon is created in the cavity for every electron that transits
through the dot contributing to the net spin current. In the
steady state, the rate at which photons are lost from the cavity
must exactly balance the rate at which photons are created by
electron spin flips in the dot. As a result, it would be possible
to measure the spin current, which is usually a difficult task, by
measuring the photo-current or, to measure the creation of one
photon states inside the cavity by measuring the charge state of
the dot, which can be done with an adjacent quantum point contact
\cite{Wei}.

\begin{figure}[htb]
\includegraphics[height=2.7in,width=3.4in]{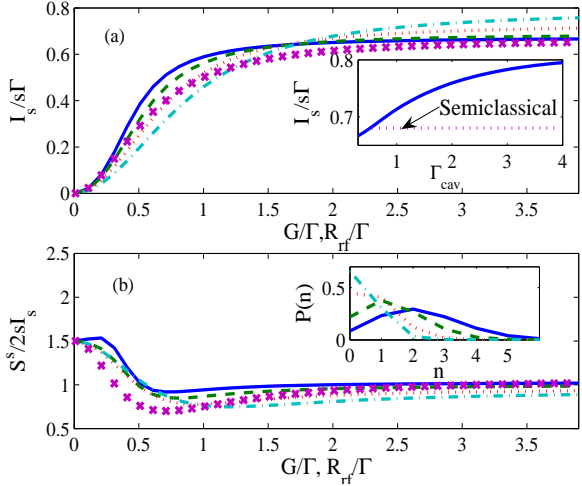}
\caption{(a) Spin current and (b) zero frequency shot noise vs.
Rabi frequency calculated for different cavity decay rates: solid
line ($\Gamma_{cav}=0.3\Gamma$), dashed line
($\Gamma_{cav}=0.5\Gamma$), dotted line ($\Gamma_{cav}=\Gamma$),
dashed-dot line ($\Gamma_{cav}=2\Gamma$), and 'x's are for the
classical e. m. field. Inset of fig. 2(a) is spin current vs.
cavity decay rate in units of $\Gamma$ and inset of fig. 2(b) is
the photon probability distribution for $G=4\Gamma$.}\label{FIG.1}
\end{figure}

In Fig. $3 (a)$ we compare the spin current in our system with the
system studied in Ref. \onlinecite{dong} where a classical e.m.
field was used to drive the transition between the different spin
states. This semi-classical model can be obtained from our quantum
model by making the replacement $G\hat{a}\rightarrow R_{rf}$,
where $R_{rf}$ is the classical field Rabi frequency, and by
eliminating all cavity degrees of freedom from the density
operator, i.e. $\rho^{(n,m)}_{\sigma,\sigma'}\rightarrow
\rho_{\sigma,\sigma'}$ and
$\dot{\rho}^{(n,m)}_{\sigma,\sigma'}|_{cavity}=0$. The
semiclassical stationary solution for the spin current is
$I_{s}=2sR_{rf}^2\Gamma/(\Gamma^{2}+3R_{rf}^{2})$. For
$R_{rf}>\Gamma$ the populations saturate,
$\bar{\rho}_{0,0}=\bar{\rho}_{\downarrow,\downarrow}=\bar{\rho}_{\uparrow,\uparrow}=1/3$,
leading to a maximum spin current of $2s\Gamma/3$. However, this
is not the case with the quantum field for $G>\Gamma$ where the
stationary populations are no longer equal,
$\bar{\rho}_{0,0}=\bar{\rho}_{\downarrow,\downarrow}>\bar{\rho}_{\uparrow,\uparrow}$,
which leads to a current that saturates at a value {\em greater}
than the semiclassical value of $2s\Gamma/3$ as shown in the inset
of Fig. $3(a)$.

The population imbalance created when $G>\Gamma$,
$\bar{\rho}_{\downarrow,\downarrow}-\bar{\rho}_{\uparrow,\uparrow}>0$,
is different from what one would expect from laser theory
\cite{meystre}. For $G<\Gamma$, a 'lasing' population inversion,
$\bar{\rho}_{\downarrow,\downarrow}-\bar{\rho}_{\uparrow,\uparrow}<0$,
does occur. The anamolous population difference is most easily
explained for $\Gamma_{cav}\gg \Gamma$ since for increasing
$\Gamma_{cav}$ the anamolous population imbalance increases (inset
of Fig. 3(a)). In this limit only the states $|0,0\rangle$,
$|\uparrow,0\rangle$, $|\downarrow,1\rangle$ and
$|\downarrow,0\rangle$ have non-negligible populations. When a
spin up electron enters the dot, it undergoes a spin flip creating
the state $|\downarrow,1\rangle$ in a time $G^{-1} \ll
\Gamma^{-1}$, which then quickly decays to  $|\downarrow,0\rangle$
in a time $\Gamma_{cav}^{-1}\ll \Gamma ^{-1}$. The spin down
electron then remains in the dot for a time $\sim \Gamma^{-1}$
after the photon has left the cavity. This time is much longer
than all other time scales and represents a bottleneck preventing
the creation of more photons. This gives rise to the larger
population in $|\downarrow,0\rangle$ compared to
$|\uparrow,0\rangle$. This is similar to the single atom laser
where the photo-current saturated as a result of the finite time
it took to recycle population in the atom \cite{mckeever}. By
contrast, when $\Gamma_{cav}\gg \Gamma > G$, the time spent in the
state $|\uparrow,0\rangle$ is longer than the lifetime of
$|\downarrow,0\rangle$. In this case one has
$\bar{\rho}_{\uparrow,\uparrow}>\bar{\rho}_{\downarrow,\downarrow}$.

Additional information can be obtained by examining the shot noise
for the spin current and the photon current. The noise power
spectrum for the current can be expressed as the Fourier transform
of the current-current correlation function,
\begin{equation}
S_{I_{\nu,\sigma}I_{\nu',\sigma'}}(\omega)=2\int_{-\infty}^{\infty}dt
e^{i \omega t}[\langle I_{\nu,\sigma}(t)I_{\nu',\sigma'}(0)\rangle
- \langle I_{\nu,\sigma} \rangle \langle I_{\nu',\sigma'} \rangle
].
\end{equation}
For the symmetric coupling, which we consider in this paper,
$S_{I_{L,\sigma}I_{L,\sigma'}}(\omega)=S_{I_{R,\sigma}I_{R,\sigma'}}(\omega)=
2I_{\sigma}\delta_{\sigma,\sigma'}+S_{I_{L,\sigma}I_{R,\sigma'}}(\omega)=S_{\sigma,\sigma'}$
where the $2I_{\sigma}\delta_{\sigma,\sigma'}$ term is the
classical Schottky noise \cite{ivana,Blanter}. The spin current
shot noise is defined as
\begin{equation}
S^{(s)}=s^2(S_{\uparrow,\uparrow}+S_{\downarrow,\downarrow}-S_{\uparrow,\downarrow}-S_{\downarrow,\uparrow}).
\end{equation}
The photo-current shot noise, $S^{(ph)}$, is defined the same way
as $S_{I_{\nu,\sigma}I_{\nu',\sigma'}}(\omega)$ with the
substitution $I_{\nu,\sigma}\rightarrow I_{photon}$. Here we use
the approach we developed in Ref. \onlinecite{ivana} to evaluate
the shot noise.

The zero frequency spin shot noise as a function of the Rabi
frequency behaves similarly for both the quantum and classical
fields (Fig. 3(b)). In the semiclassical case, we obtained for the
zero frequency noise
\begin{equation}
S^{(s)}/2sI_{s}=(3\Gamma^{4}+2\Gamma^{2}R_{rf}^{2}+19R_{rf}^{4})/2(\Gamma^{2}+3R_{rf}^{2})^{2}.
\end{equation}
For weak coupling, $G,R_{rf}\ll\Gamma$, the shot noise is
super-Poissonian, approaching $3/2$ as $G,R_{rf}\rightarrow 0$.
Increasing the dot-field coupling decreases the noise until the
noise plataeus for $G,R_{rf}\gg\Gamma$. The plateau value is
$\simeq 2sI_s$ for the classical field but for the quantum field
it becomes increasingly sub-Poissonian for increasing
$\Gamma_{cav}$.


\begin{figure}[htb]
\includegraphics[height=2.2in,width=3.5in]{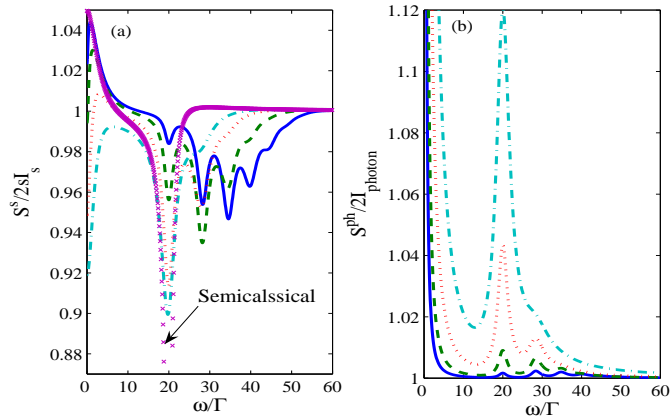}
\caption{The frequency dependent spin current (a) and
photo-current (b) shot noise vs. frequency for $R_{rf}=G=10\Gamma$
and different $\Gamma_{cav}$(curve labels same as Fig. 2).
}\label{FIG.2}
\end{figure}

The frequency dependent shot noise provides even more information
as can be seen in Fig. 4. For strong coupling, $G\gg \Gamma$,
$\Gamma_{cav}$, dips (peaks) are present in the spin (photon)
current noise spectrum, which are located precisely at the cavity
Rabi frequencies $2G\sqrt{n+1}$ for $n=0,1,2,...$. By comparison
$S^{(s)}(\omega)$ exhibits a single dip at $\omega=2R_{rf}$ for
the semiclassical system. These resonances in the noise spectrum
represent a definite sign of the quantization of the cavity field.
For $\Gamma_{cav}<\Gamma$ there is a large probability that more
than one electron will tunnel through the dot before the photons
start to leave the cavity. Whenever one electron passes through
the dot, the number of photons increases by one. This process
continues until the cavity starts to decay. For $\Gamma_{cav}\ll
\Gamma$, this will lead to a non-negligible probability for the
cavity to be in a state with $n=0,1,2,3...$ photons.
Qualitatively, each one of these photon states will contribute to
the shot noise spectrum with a weight $P(n)$, where $P(n)$ is the
probability of having $n$ photons, which is shown in the inset of
Fig. 3(b). By increasing $\Gamma_{cav}$, the number of populated
photon states decreases until finally for $\Gamma_{cav}\gg
\Gamma$, only the dip (peak) at the vacuum Rabi frequency, $2G$,
remains in the spectrum. We have fitted these dips to Lorentzians
and found them to have a width of $\Gamma+(n+1)\Gamma_{cav}/2$ for
$\Gamma_{cav}\leq\Gamma$. Decreasing $G$ will decrease the
separation between the dips (peaks) until they begin to overlap
when $G \lesssim \Gamma, \Gamma_{cav}$.

\section{Conclusion}
In conclusion, we have studied the spin current and spin current
shot noise generated by a quantum dot coupled to single mode of an
optical microcavity. We found that the spin current can be
significantly larger than for a classical driving field due to
cavity decay of the quantized field. Most importantly, we showed
that the spin current shot noise exhibits clear signatures of the
discrete nature of the photon states in the cavity in the limit of
strong cavity coupling.

Although no experiments have yet been performed studying transport
through quantum dots embedded in cavities, it is very likely that
the two avenues of research that have been pursued with
self-assembled dots namely transport \cite{schmidt,ota,barthold}
and cavity-QED \cite{reithmaier,peter} will intersect in the near
future. The main experimental challenge to such experiments are
connecting metal electrodes to the devices, which are optically
very absorbing and would lead to a strong reduction in the
Q-factor of the cavity \cite{strauf}. This might be ameliorated by
using electrodes consisting of doped semiconductors with carrier
concentrations below that of metals. On the other hand, the
enhancement of the spin current over the semiclassical limit
should be observable even with a low-Q cavity.



\begin{references}
\bibitem{note} We use the level structure defined in Ref. [11] rather
than the conventions used in Refs. [9] and [14] where the spin
down state has lower energy than the spin up state.
\end{references}
\end{document}